\documentclass[aps,prd,preprint,superscriptaddress,tightenlines,%
nofootinbib]{revtex4}
\newcommand{\PRE}[1]{{#1}}   

\usepackage{bm}
\usepackage{epsfig}

\newcommand{\eqref}[1]{Eq.~(\ref{#1})}

\begin{document}

\title{
\PRE{\vspace*{.9in}}
{\Large {\bf The Pierre Auger Observatory}}  
\PRE{\vspace*{0.3in}}
}
\author{{\bf John Swain, for the Pierre Auger Observatory}\\ 
{\it Department of Physics,
Northeastern University,
Boston, MA 02115}\\
{\tt john.swain@cern.ch}
\PRE{\vspace*{.1in}}
}

\begin{abstract}
\PRE{\vspace*{.3in}}
One of the most fascinating puzzles in particle 
astrophysics today is that of the origin and nature
of the highest energy cosmic rays.
The Pierre Auger Observatory (PAO), currently
under construction in Province of Mendoza, Argentina, and with another
site planned in the Northern hemisphere, is a major
international effort to make precise, high statistics
studies of the highest energy cosmic rays.
It is the first experiment designed to work in a
hybrid  mode incorporating both a ground-based
 array of 1600 particle detectors
spread over 3000 km$^2$ with fluorescence telescopes placed on the
boundaries of the surface array. The 
current status of the observatory is presented
and prospects for the future discussed.

\PRE{\vspace*{.9in}}

\end{abstract}

\maketitle

\section{Introduction}

The observation of ultra high energy
cosmic rays (UHECR) with energies above $10^{20}$ eV \cite{ev},
is puzzling in at least two ways. 
First of all, it is quite
difficult to conceive of acceleration mechanisms which are adequate
to impart such enormous energies to cosmic ray particles\cite{bh,ol} --
energies comparable to those carried by everyday objects like tennis
balls or golf balls! 
Secondly, even if such a mechanism is found, it is difficult to see how
such high energy particles would
make it through the background radiation: the celebrated GZK cutoff\cite{gzk}
predicts
that protons over about $5 \times 10^{19}$ eV should rapidly lose energy
in inelastic collisions with the cosmic microwave background photons
with similar energy degradation mechanisms being present for most other
particles, including heavy nuclei.

A very  important point to make early in this talk is that while there is
a puzzle concerning so-called ``super-GZK events'', the Pierre Auger
Obseratory will provide interesting information regardless of how well
the GZK cutoff holds out.  If there is no GZK cutoff, this will be
an unambiguous sign of new physics.
If the GZK cutoff
is found, then we can rest assured of that piece of the physics and
confidently use the data to try to understand the nature of the sources.
This point is often missed by people who see the whole physics motivation
as ``is there a GZK cutoff or not?''.


Above $10^{15}$ eV, cosmic ray primaries are not
detected directly, but rather through 
the effects that such particles produce when they strike the upper atmosphere. 
There they initiate a cascade of reactions, some nuclear, but most forming an
electromagnetic shower made of repeated bremsstrahlung and $e^+e^-$ pair
creation events. This shower can be detected experimentally 
through the fluorescence it produces in the atmosphere ( due to excited nitrogen)
or via the particles that reach the ground.


Our understanding of the highest energy part of the
spectrum above the so-called ``ankle'' ($5 \times 10^{18}$ eV) 
is poor due to a combination of
low statistics, uncertain energy resolution,
uncertainties in energy conversion arising from models, and a 
lack of knowledge of the mass composition and the fluorescence
yield efficiency.
Clearly there is something interesting going on, but to fully understand the
situation we need more statistics, better control over systematic uncertainties,
and full sky coverage: enter the Pierre Auger Obervatory.
A comprehensive review of the state of the art prior to the Pierre Auger
Observatory can be found in \cite{bigreview}. 

\section{The Pierre Auger Observatory}

The Pierre Auger Observatory (PAO) is actually comprised of two sub-observatories,
one currently under construction in Mendoza, Argentina since 2000.
This site is especially 
interesting, since in addition to being in the wine-growing district, it
offers a view of the centre of the galaxy.
Another
is planned for the Northern hemisphere, and while for the remainder
of this talk I will
concentrate on the Southern site it is important to understand that the full
observatory is comprised of two sites. This will eventually allow full-sky
coverage which is very important to allow good studies on
anisotropies\cite{anisotropies}.

The PAO is designed to measure the energy, 
arrival direction and  primary species with excellent
precision and very high statistics. A unique feature of the design is the
combination of both fluorescence detection and ground-based particle
detectors which can be operated independently as well as together in 
``hybrid'' mode.

The scale of the observatory was determined by the requirement that
we can collect high 
statistics in and around the region of the expected GZK cutoff, with 1600 particle detectors
separated from each other by 1.5 km and 
covering an area of 3000 km$^2$, overlooked by four fluorescence detectors
which can only operate when ambient conditions offer a clear, dark sky, 
which leads to a roughly 10\% duty cycle. Figure 1 shows, together with one
of the surface detectors, a photograph
of one the fluorescence detectors where both the mirror and the box of 440 photomultipliers
which register the fluorescence light can be seen. The fluorescence measurements
are complemented by a a very comprehensive atmospheric monitoring system.
The surface array stations are water \v{C}erenkov detectors and can operate
continuously.

The fact that about 10\% of the showers detected by PAO 
will be observed by {\em both} surface and fluorescence  detectors, offers
the possibility of doing calibrations and understanding systematic errors
in a manner that has never been possible before.
Access to a large-dimensional parameter 
space of observables should allow not only determination of the
direction and energy of incoming primaries, but also the disentanglement
of information about composition from the notoriously difficult 
systematic errors associated with the choice of hadronic interaction
models.

Each ground-based detector is a cylindrical, opaque  tank of 10 m$^2$ and a
water depth of 1.2 m, where particles  produce light by \v{C}erenkov radiation.
The filtered water is contained in a bag which diffusely
reflects the light collected by three photomultipliers (PMT's) installed on the
top. The large diameter PMT's ($\approx$ 20 cm ) 
are mounted facing down and look at the water through sealed polyethylene 
windows that are integral part of the internal liner. The signals are processed
locally and a second level trigger is identified  before transmitting the data
to the central acquisition system  \cite{Tiina}. The fact that the tanks
are quite
deep enables showers to be detected efficiently over a wide angular range. Due
to the size of the array, the stations have to be able
to function independently and yet in communication with the central data
acquisition system.
The stations operate on battery-backed solar power  and  communicate with a
central station by using wireless LAN radio links \cite{Paul}. Absolute timing
information is obtained from the Global Positioning System (GPS)
\cite{clejer} and is used to reconstruct 
the direction from which the primary came.
 Figure 1 shows a water \v{C}erenkov detector installed  in the
Southern Observatory as well as one of the fluorescence detectors.
Mounted on top of the tank are  the solar panel,
electronic enclosure, mast, radio antenna and GPS antenna. The
battery is housed in a box attached to the tank.

\begin{figure}[htbp]
\hbox{\epsfig{file=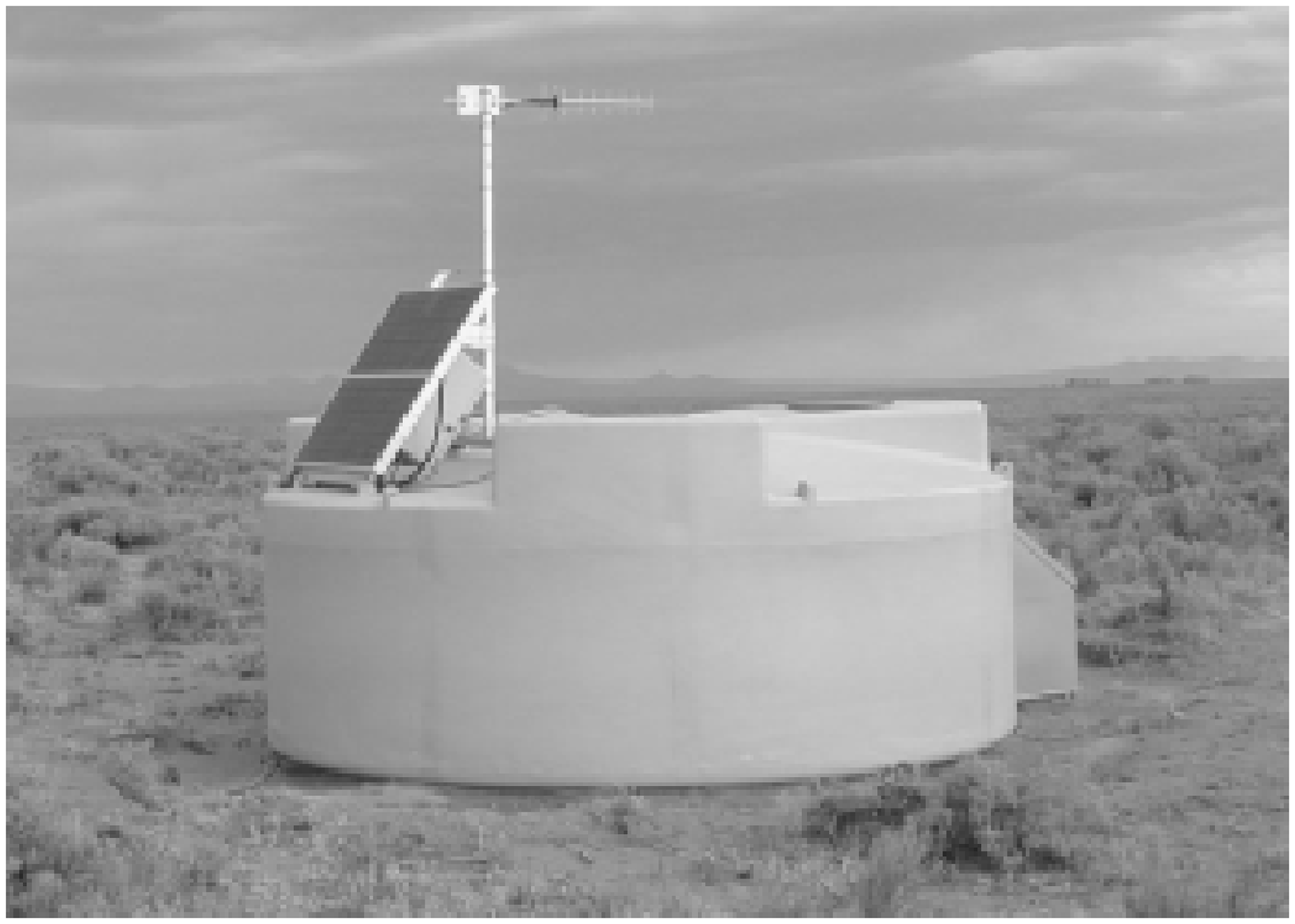,width=0.60\textwidth,clip=}
\epsfig{file=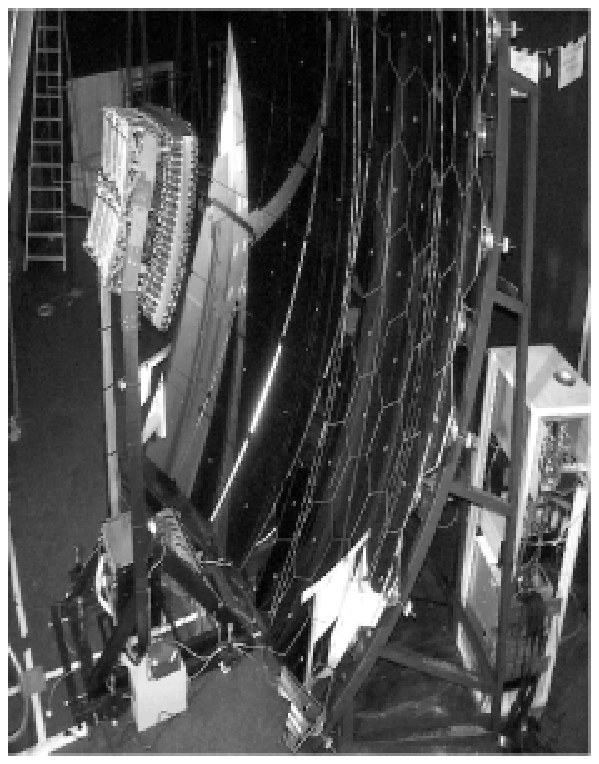,width=0.35\textwidth,clip=}}
\caption{\label{detectors} Photographs of a typical surface detector and one of the
fluoresence detectors showing the mirrors and the array of phototubes onto which
they direct the collected light.}
\end{figure}

The expected angular resolution for the ground array of the Southern Auger
Observatory is less than 1$^\circ$ for all energies,
and better for
large events above  $ 10^{20} $eV. The expected energy resolution is estimated
to be 12\%, averaged over all energies (assuming a proton-iron primary
mixture), falling to 10\% at  $  10^{20} $eV.  The limiting aperture for the
full Southern Observatory array and for zenith angle less than  60$^\circ$ is 
7350 km$^2$sr. The  detection efficiency at the trigger level should reach
100\% for energies above  $ 10^{19} $eV \cite{Jere}. Additionally, if events
above 60 degrees can be analyzed effectively, the aperture will increase by
about 50\% .

In hybrid mode, the Pierre Auger Observatory is expected to have 6\%
energy  resolution and an angular precision of 0.5$^\circ$ at $ 10^{20} $eV
where only statistical errors are taken into account in these estimates.
The detector is optimized for energies above  $10^{19} $eV, with good
reconstruction expected at energies down to 1 EeV. The hybrid data set will
provide the best evaluation of primary species, allowing a simultaneous fit to
all parameters sensitive to mass composition.
%

The first cosmic ray event  detected by one of the two prototype telescopes
installed at Los Leones is displayed in Figure 2. A twenty pixel track, produced
by light from a shower, with a length of 8 $\mu $s can be seen. The angular
velocity of the shower image across the sky allows the distance of the shower core
to be established as 5 km. The time duration of the signal in the field of view of
the telescope corresponds to a track length of 2.4 km. The mirror inverts the
image, and particles from the sky appear to be going upwards as seen by
the camera.

\begin{figure}[htbp]
\epsfig{file=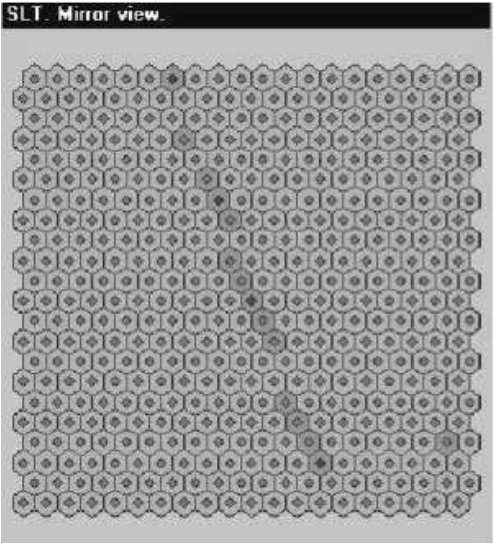,width=0.4\textwidth,clip=}
\caption{First high energy cosmic ray observed by one of the prototype
telescopes at Los Leones.}
\end{figure}

Construction continues, and we look forward to the large amount
of very high quality which will appear in the years to come, and
the light it will shed on one of nature's great mysteries.

\section*{Acknowledgments}
I would like to thank the organizers of this conference for a most
interesting meeting, as well as all my collaborators in the Pierre Auger
Observatory. Special thanks are due to Maria Teresa Dova for assistance
in the preparation of this talk, and, as always, I would like to thank
the National Science Foundation for its continued support.

\end{document}